\documentclass[%
 aip,
 amsmath,amssymb,
 reprint,%
]{revtex4-1}

\usepackage{graphicx}
\usepackage{dcolumn}
\usepackage{bm}

\usepackage[utf8]{inputenc}
\usepackage[T1]{fontenc}
\usepackage{mathptmx}
\usepackage{etoolbox}
\usepackage{placeins}
\usepackage{xcolor}
\usepackage{color}
\usepackage{soul}

\makeatletter
\def\@email#1#2{%
 \endgroup
 \patchcmd{\titleblock@produce}
  {\frontmatter@RRAPformat}
  {\frontmatter@RRAPformat{\produce@RRAP{*#1\href{mailto:#2}{#2}}}\frontmatter@RRAPformat}
  {}{}
}%
\makeatother

\begin{document}

\preprint{AIP/123-QED}

\title[ Visibility function and Stokes parameter analysis for non-ideal quarter-wave plates]{Visibility function and Stokes parameter analysis\\ for non-ideal quarter-wave plates in a polarization contrast scheme}

\author{ S.\,A.~Kupavskii$^*$}
\affiliation{%
 Lomonosov Moscow State University, Faculty of Physics, Moscow 119991, Russia.
}%

\author{R.\,O.~Malikov}
\affiliation{%
 Lomonosov Moscow State University, Faculty of Physics, Moscow 119991, Russia.
}%
\affiliation{Prokhorov General Physics Institute of the Russian Academy of Sciences, Moscow 119991, Russia} 
\affiliation{Skolkovo Institute of Science and Technology, Moscow 143025, Russia}

\author{S.\,A.~Dyakov}%
\affiliation{Skolkovo Institute of Science and Technology, Moscow 143025, Russia}
\author{S.\,G.~Tikhodeev}

\affiliation{%
 Lomonosov Moscow State University, Faculty of Physics, Moscow 119991, Russia.
}%
\affiliation{Prokhorov General Physics Institute of the Russian Academy of Sciences, Moscow 119991, Russia} 
\affiliation{Skolkovo Institute of Science and Technology, Moscow 143025, Russia}

\author{A.\,A.~Gavrilov}%
\affiliation{
Institute of Solid State Physics, Russian Academy of Sciences, Chernogolovka 142432, Moscow region, Russia
}%

\author{K.\,R.~Dzhikirba}%
\affiliation{
Institute of Solid State Physics, Russian Academy of Sciences, Chernogolovka 142432, Moscow region, Russia
}%

\author{S.\,A.~Andreeva}%
\affiliation{National Research University Higher School of Economics, Moscow 101000, Russia}
\affiliation{
Institute of Solid State Physics, Russian Academy of Sciences, Chernogolovka 142432, Moscow region, Russia
}%

\author{I.\,V.~Kukushkin}%
\affiliation{
Institute of Solid State Physics, Russian Academy of Sciences, Chernogolovka 142432, Moscow region, Russia
}%

\author{N.\,A.~Gippius}%
\affiliation{Skolkovo Institute of Science and Technology, Moscow 143025, Russia}

\email{kupavskiysa@my.msu.ru}

\date{\today}

\begin{abstract}
The match of the fourth power of the normalized third Stokes parameter for a single non-ideal quarter-wave plate (QWP) and the visibility function of the polarization contrast measured using a polarimetric setup with parallel polarizers and two identical QWPs was examined and observed. 
 This match was first established numerically
 using the Fourier-modal method (FMM) and described within the Jones matrix formalism. 
An analytical expression for the visibility function of the polarization contrast scheme in terms of the normalized Stokes parameters was then derived, along with the asymptotic expansions in terms of small parameters representing the phase deviation, transmission anisotropy, and angular misalignment of optical components of the polarization contrast scheme from their ideal orientations.
For the investigated silicon-grating-based QWP, good quantitative agreement between the FMM-calculated fourth power of the normalized third Stokes parameter, 
and the visibility function (FMM-calculated and experimentally measured) was demonstrated 
over almost the entire investigated frequency range of 200--700 GHz.
The established match paves the way for a rapid, simple and broadband characterization of the QWPs quality.
\end{abstract}

\maketitle

Polarization-sensitive measurements are crucial for analyzing materials with birefringence, optical anisotropy, and chirality~\cite{Grischkowsky1990,Jepsen2011,Katletz2012,Choi2019}. The polarization contrast method~\cite{Goldstein2011,Collett2005} is a particularly effective polarimetric setup for this analysis. This method uses a visibility function to measure the contrast between two configurations -- parallel and perpendicular -- of identical quarter-wave plates (QWPs). When the visibility function reaches values close to one, it signifies the highest quality of these QWPs. Another important characteristic of QWP is the normalized third Stokes parameter, which quantifies the degree of circular polarization of light passing through a QWP. For high-quality QWPs, this parameter approaches unity. 

In our investigation, we establish that the visibility function within the polarization contrast method and the fourth power of the normalized third Stokes parameter of a non-ideal QWP exhibit similar frequency behavior across a wide spectral range near the peaks of the visibility function. This correlation between the two measures enables straightforward conversion, which can be practically applied for swift and extensive evaluation of the QWP's optical quality. Since the polarization contrast method does not require the use of expensive equipment -- only linearly polarizing filters -- the correlation identified will enable rapid quality control and broadband characterization of QWPs, e.g., during their manufacturing.

In this work, an exact expression for visibility in terms of Stokes parameters is derived using the Jones matrix formalism, along with an asymptotic expansion in two small parameters:  the deviation of the QWP phase from $\pi/2$ and the asymmetry of the optical transmission coefficients around the quarter-wave points. The identical second-order asymptotic behavior of the visibility function and the fourth power of the normalized third Stokes parameter is  established.

In addition to these two aspects, the effect of angular misalignment from ideal orientations of optical devices is considered. In particular, asymptotic expansions for the visibility function and the fourth power of the normalized third Stokes parameter are obtained. Accounting for the angular misalignment enables the experimentally measured visibility spectrum to be described theoretically across almost the entire measurement range of 200--700 GHz for the investigated silicon grating-based QWPs. Numerical calculations confirm the experimental and analytical results.

Note that in the Jones matrix formalism, multiple coherent scattering between QWPs as well as between QWPs and linear filters is neglected. This can be done because i) the reflectivity of the optimized optical elements is small and ii) the distance between them is large compared with the coherence length of the electromagnetic wave.

\begin{figure*}
    \includegraphics[width=\textwidth]{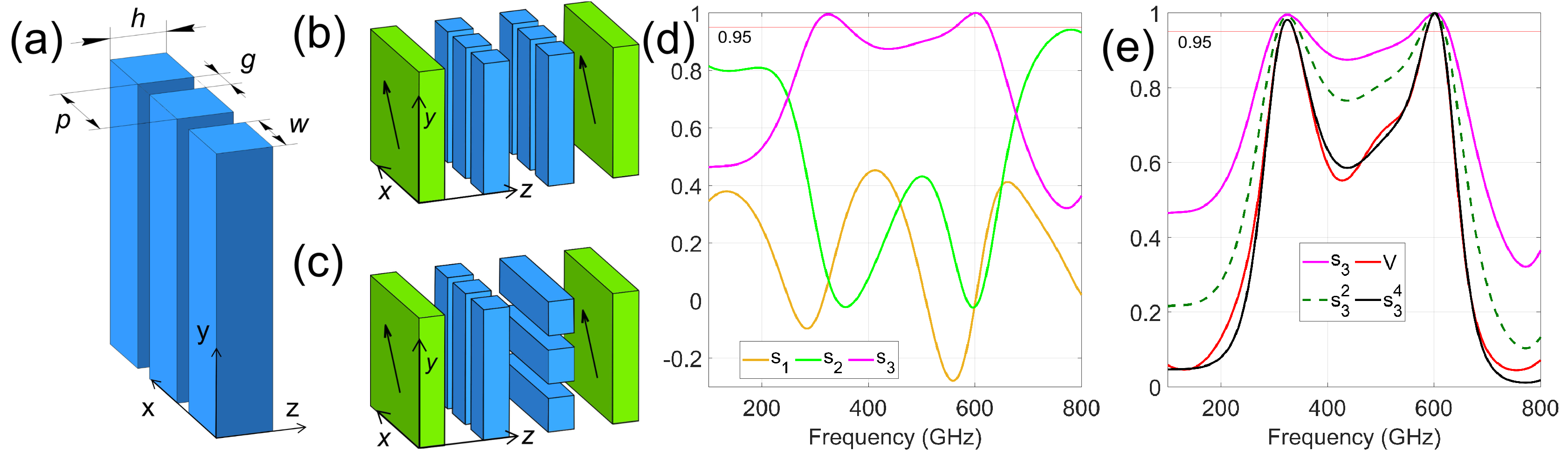}
    \caption{\label{fig:fig1} (a)~Geometry of QWP (all sizes in $\mu$m): grating with a height $h=172$, a period of grating $p=140$, a grating width $w=103$, air gap width $g=37$, material is silicon with dielectric permittivity $\varepsilon=11.6622$; (b and c)~Polarization contrast scheme for parallel and perpendicular QWPs, respectively; (d)~FMM-calculated spectra of the normalized Stokes parameters for a single QWP; (e) FMM-calculated spectra of the normalized third Stokes parameter, its second and fourth powers and the visibility function $V$ for the  polarization contrast scheme. Red horizontal line denotes value of 0.95.} 
\end{figure*}

The effect under consideration was observed when characterizing a terahertz QWP based on a silicon grating~\cite{Zhang2015}, the geometric parameters of which were optimized for two frequencies: $330$~GHz and $600$~GHz. This one-dimensional grating has a height $h=172 \, \mu$m, a grating period  $p=140 \, \mu $m, a grating width $w=103 \, \mu $m, and an air gap width $g=37 \, \mu $m, the material is silicon with dielectric permittivity $\varepsilon=11.6622$ (Fig.~\ref{fig:fig1}(a)).

To characterize such a QWP, a polarization contrast measurement scheme was used, with identical parallel (and perpendicular) QWPs between the co-oriented linear polarizer and analyzer, as shown in Fig.~\ref{fig:fig1}(b) and (c), respectively. Here, the angle between the QWPs and the analyzer and polarizer is equal to $45^\circ$. 

The visibility function $V$ in this scheme is defined as 
\begin{equation}
\label{eq:polarizCOntrast}
    V = \frac{I_\perp - I_\parallel}{I_\perp + I_\parallel},
\end{equation}
where $I_\parallel$ and $I_\perp$ are the intensities of transmitted light with parallel and crossed QWPs, respectively.

The structure examined in this study is not a broadband QWP. To illustrate the relevance of the correlation discussed here between visibility and the fourth power of the normalized third Stokes parameter, $V \approx s_3^4$, in assessing the quality of quarter-wave plates across a broad frequency spectrum, we deliberately chose a structure where the normalized third Stokes parameter surpasses 0.95 in two narrow bands around 330 and 600 GHz. These are separated by a wide region where $s_3 < 0.95$ (indicated by horizontal red lines in Figs.~\ref{fig:fig1}(e, d)), rendering the QWP inadequate. We plan to discuss the possibilities to improve the quality of such QWPs elsewhere.

The structure under consideration is a one-dimensional photonic crystal slab, and the Fourier-modal method (FMM) in the scattering-matrix form~\cite{Whittaker1999, Tikhodeev2002, Gippius2005} was used to calculate the electromagnetic fields and optical coefficients of reflection and transmission in the systems. Then, the function $V$ of the polarization contrast scheme and the Stokes parameters of a single QWP were calculated. The results on the basis of the p(x)- and s(y)-polarization for the normalized Stokes parameters
$s_1=S_1/S_0$, $s_2 = S_2/S_0$, $s_3=S_3/S_0$ (where 
 $S_0 = |E_x|^2 +|E_y|^2$, $S_1=|E_x|^2-|E_y|^2$, $S_2 = 2 \text{Re} (E_x E_y^*)$, $S_3 = -2 \text{Im}  (E_x E_y^*)$)
are presented in Fig.~\ref{fig:fig1}(d). Here  $\mathbf{E}=\begin{pmatrix} E_x & E_y \end{pmatrix}^T$ is the electric field vector passing through a single oriented QWP as shown in Fig.~\ref{fig:fig1}(a). Calculations using FMM were performed neglecting multiple scattering between different optical elements as in the Jones matrix formalism.

Figure~\ref{fig:fig1}(e) compares the FMM-calculated visibility spectra  $V$, Eq.~(\ref{eq:polarizCOntrast}), with $s_3$,  $s_3^2$ and $s_3^4$. As we can see, the fourth power of the normalized third Stokes parameter $s_3^4$ nicely coincides with the visibility function $V$ in the whole investigated frequency range.

The reason for this can be analytically understood using the Jones matrix formalism. The electric field vector of the incident wave has the form $\mathbf{E}_0 = 1/\sqrt{2} \, \begin{pmatrix} 1 & 1\end{pmatrix}^T$ on the  basis of p(x)- and s(y)-polarizations ($\mathbf{E}_0 = \begin{pmatrix} 0 & 1\end{pmatrix}^T$ in the coordinate system of the polarizer and analyzer aligned along the arrows in Fig.~\ref{fig:fig1}(b, c)). Then, the electric vector propagates through the parallel (Fig.~\ref{fig:fig1}(b))  or crossed (Fig.~\ref{fig:fig1}(c)) arrangement of two QWPs, both of which are rotated by $\pm 45^\circ$ with respect to the polarizer and analyzer. The polarizer and analyzer are described by the Jones matrix as
\begin{equation}
\label{eq:polarizMatrix}
    \mathcal{P} \equiv \mathcal{A} = \dfrac{1}{2} \begin{pmatrix} 1 & 1 \\ 1 & 1\end{pmatrix}.
\end{equation}
The Jones matrices of the first QWP (as well as of the second one in the parallel configuration) and of the second QWP in the crossed configuration are, respectively,
\begin{equation}
\label{eq:DGmatrix}
\mathcal{M} =
\begin{pmatrix}
t_{p} & 0 \\
0 & t_{s} 
\end{pmatrix}, \,
\mathcal{M}_{\perp} = 
\begin{pmatrix}
t_{s} & 0 \\
0 & t_{p} 
\end{pmatrix},
\end{equation}
where $t_s$ and $t_p$ are complex amplitude transmission coefficients in s- and p-polarizations, respectively. Thus, the outgoing electric field vectors after transmitting through these configurations take the form
\begin{equation}
    \mathbf{E}_{\parallel} = \mathcal{A} \mathcal{M}^2 \mathcal{P} \mathbf{E}_0, \, \mathbf{E}_{\perp} = \mathcal{A} \mathcal{M}_{\perp}\mathcal{M} \mathcal{P} \mathbf{E}_0,
\end{equation}
and their intensities are $I_{\parallel,\perp}=|\mathbf{E}_{\parallel,\perp}|^2$.
Note here that Eq.~(\ref{eq:DGmatrix}) holds if the linear p- and s-polarizations are decoupled. This works for the QWPs considered here based on the 1D photonic crystal slabs with normal light incidence. In a more general case, the Jones matrices (\ref{eq:DGmatrix}) also contain non-diagonal terms, which we currently neglect.

The outgoing electric field vector, after transmitting through a single QWP, has the form
\begin{equation}
    \mathbf{E}=\mathcal{M}\mathbf{E}_0.
\end{equation}

The symmetry of the QWP in question is D$_{\text{2h}}$ and, generally, the complex amplitude transmission coefficients of the QWP are non-equal, $t_s \neq t_p$. (For QWPs made of ordinary birefringent material, anisotropy of $\kappa$ causes the same effect~\cite{Grischkowsky1990,Jepsen2011}.)

Let $t_{p,s}=\sqrt{T_{p,s}}\text{exp}(i\Delta_{p,s})$, where $\Delta_{p,s}$ are the phases of $t_{p,s}$ and $T_{p,s}$ are the intensity transmission coefficients. Introducing a QWP phase shift $\delta=\Delta_s-\Delta_p$ (which in the quarter-wave point has to be equal to $\pi/2$) we can obtain the visibility function in the polarization contrast scheme as
\begin{equation}
\label{eq:V_dichro}
    V = \dfrac{4\,T_p T_s\,\sin^2\!\delta \;-\; (T_p-T_s)^2}
    { 4\,T_p T_s\,\cos^2\!\delta + (T_p+T_s)^2}\, ,
\end{equation}
which, in terms of the normalized Stokes parameters, takes the form
\begin{equation}
     V = \dfrac{s_3^2 - s_{1}^2}{s_{2}^2+1}.
\end{equation}
The fourth power of the normalized third Stokes parameter for a single QWP takes the form 
\begin{equation}
\label{eq:Stokes_dichro}
    s_3^4 = \frac{16{T_p^2 T_s^2}\,\sin^4\delta}{(T_p+T_s)^4}.
\end{equation}
To obtain asymptotics of Eqs.~(\ref{eq:V_dichro}) and (\ref{eq:Stokes_dichro}) we define two small (in case of high-quality QWP) dimensionless parameters characterizing the deviation of the phase shift from $\pi/2$ and transmission anisotropy, respectively:
\begin{equation}
\label{eq:alpha}
    \alpha = 1 - \sin\delta,
\end{equation}
\begin{equation}
\label{eq:beta}
    \beta  = \frac{T_s}{T_p} - 1.
\end{equation}

By factoring functions $V$ and $s_3^4$ into powers of $\alpha<1$ and $|\beta|<1$ and taking terms up to the fourth order, we obtain
\begin{equation}
\label{eq:V_asym_lossy}
\begin{array}{rcl}
    V & = & 1 - 4\alpha + 10\alpha^2 - \dfrac{1}{2}\beta^2\ - 24 \alpha^3 + 2\alpha \beta^2 + \dfrac{1}{2}\beta^3 \\
    &  & + 58 \alpha^4 - 7 \alpha^2 \beta^2 - 2\alpha \beta^3 - \dfrac{3}{8}\beta^4 + O_5(\alpha,\beta),
\end{array}
\end{equation}
\begin{equation}
\label{eq:Stokes_asym_lossy}
\begin{array}{rcl}
    s_3^4 & = & 1 - 4\alpha + 6\alpha^2 - \dfrac{1}{2}\beta^2 - 4 \alpha^3 + 2\alpha \beta^2 + \dfrac{1}{2}\beta^3 \\
    &  & + \alpha^4 - 3 \alpha^2 \beta^2 - 2\alpha \beta^3 - \dfrac{5}{16}\beta^4 
+ O_5(\alpha,\beta).
\end{array}
\end{equation}

A comparison of Eqs.~(\ref{eq:V_asym_lossy}) and (\ref{eq:Stokes_asym_lossy}) shows that, indeed, the difference between the visibility and the fourth power of the third Stokes parameter begins at the second order in $\alpha, \alpha\beta$ and at the fourth order in $\beta$,
\begin{equation}
\label{eq:diver}
    V - s_3^4 = 4\alpha^2 - 20 \alpha^3 + 57 \alpha^4 - 4 \alpha^2 \beta^2 - \dfrac{1}{16} \beta^4 + O_5(\alpha, \beta).
\end{equation}

\begin{figure}[t]
    \includegraphics[width=\columnwidth]{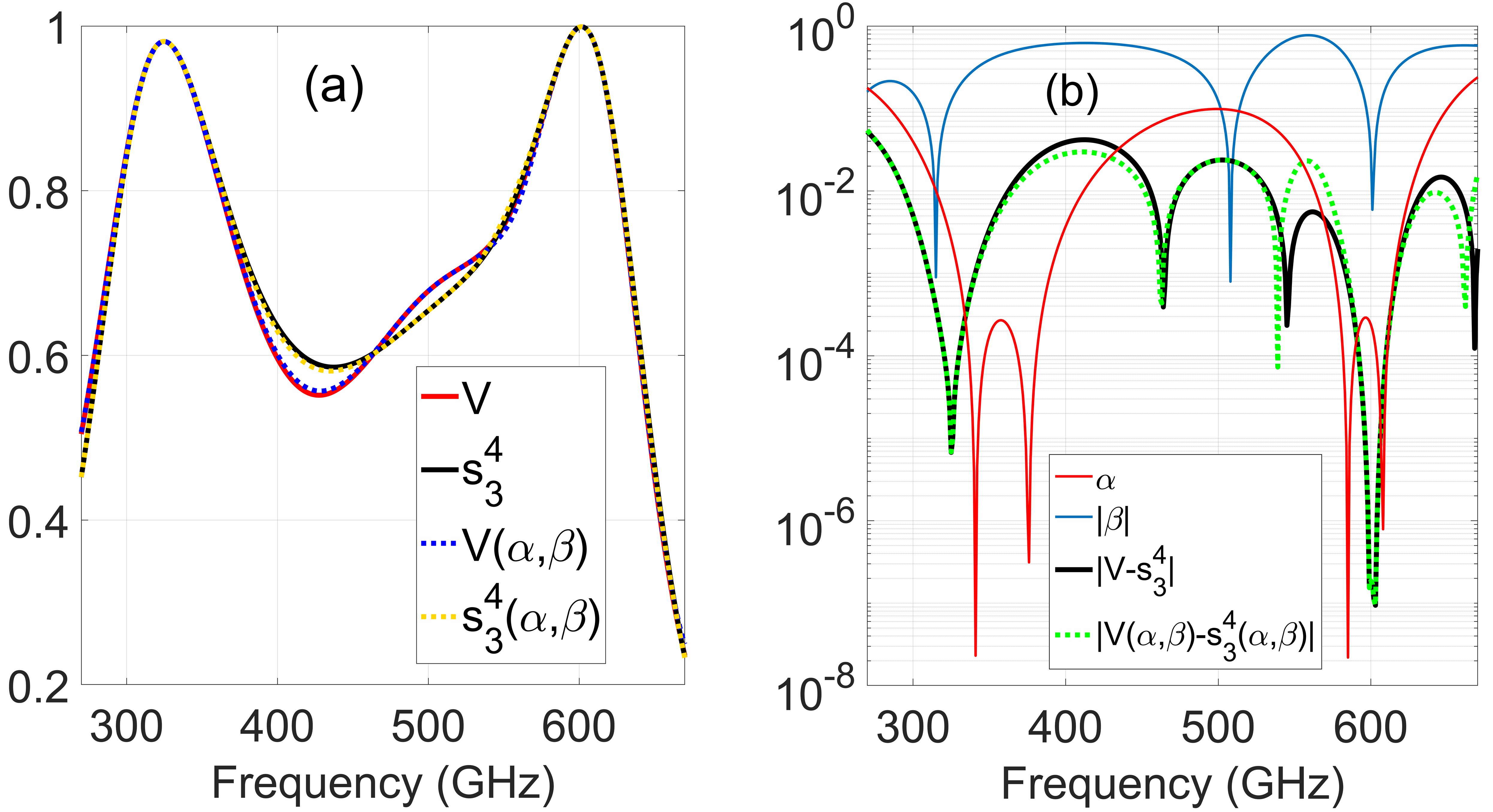}
    \caption{\label{fig:fig2}
    (a)~The frequency dependencies $V(\nu)$ and $s_3^4(\nu)$, calculated using FMM, and as approximate functions of small parameters  $\alpha(\nu)$ and $ \beta(\nu)$, extensions of Eqs.~(\ref{eq:V_asym_lossy}) and (\ref{eq:Stokes_asym_lossy}), Eqs.~(S1) and (S2). (b)~FMM-calculated $\alpha(\nu)$ and absolute value of $ \beta(\nu)$, as well the absolute values of differences between the FMM-calculated and approximate $V$ and $s_3^4$ in semi-logarithmic scale. }
\end{figure}

Figure~\ref{fig:fig2}(a) shows the frequency dependencies of the visibility function and the fourth power of the Stokes parameter, calculated using FMM, and approximate ones, calculated using extensions of Eqs.~(\ref{eq:V_asym_lossy}) and (\ref{eq:Stokes_asym_lossy}) up to the sixth and fifth order in small parameters, respectively, see Eqs.~(S1) and (S2) in the {\bf \underline{supplementary material}}. The frequency dependencies of the small parameters $\alpha$ and $|\beta|$, calculated according to the FMM by Eqs.~(\ref{eq:alpha}) and (\ref{eq:beta}), respectively, are shown in Fig.~\ref{fig:fig2}(b) on a semi-logarithmic scale. Figure~\ref{fig:fig2}(b) also shows the FMM-calculated as well the approximate absolute values of differences $V-s_3^4$ by Eqs.~(S1) and (S2). It can be seen that, indeed, the latter are small, in agreement with Eq.~(\ref{eq:diver}), provided that $\alpha$ and $\beta$ are small.

\begin{figure}
    \begin{center}
   \includegraphics[width=1\columnwidth]{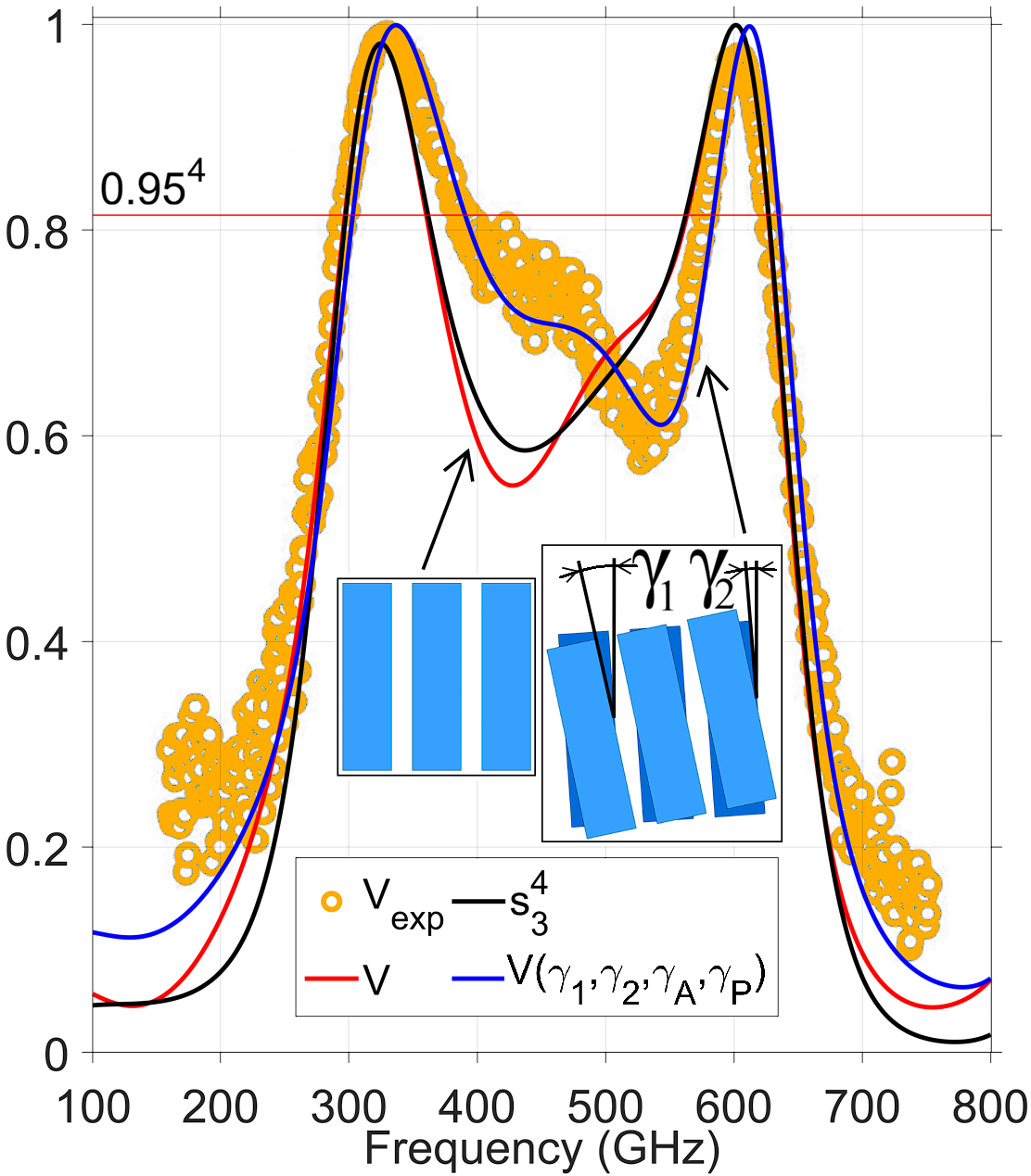}
       \end{center}
    \caption{\label{fig:fig3} 
   Experimentally measured (circles) and FMM-calculated (red solid curve) spectra of the visibility function, in comparison with the FMM-calculated spectra of $s_3^4$ (black solid curve). Blue solid curve is the FMM-calculated visibility function with accounting for misalignment, see inserts and explanation in the text. Red horizontal line marks the visibility level $0.95^4$.
    } 
\end{figure}

Let us now compare our model with the experimentally measured visibility function $V_{\text{exp}}$ for the  polarization contrast scheme described above, see the orange circles in Fig.~\ref{fig:fig3}. As can be seen, the FMM-calculated spectra of $s_3^4$ and $V$ for the ideal case of the polarization contrast scheme arrangement accurately reproduce the experimentally measured visibility function $V_{\text{exp}}$ in the regions of two peaks of the visibility function and even outside them. However, between the two peaks there is a strong mismatch. This discrepancy may be due to the fact that, in reality, when measuring $V_{\text{exp}}$, the angles between all optical devices may not be exactly maintained.

Consider the angular misalignment of optical components from their ideal orientations in the polarization contrast scheme. It appears that the visibility function is primarily influenced by $I_\parallel$, while $I_\perp$ produces minimal changes. Let $\gamma_1$ and $\gamma_2$ are small angular deviations (in radians) of the first and second QWPs in parallel configuration from the vertical orientation (see the insert in Fig.~3) and $\gamma_A$ and $\gamma_P$ are small angular deviations (in radians) of the analyzer and polarizer, respectively, from their orientations in both parallel and perpendicular configurations. Then the Jones matrices of these  elements take the forms
\begin{equation}
\begin{array}{rcl}
    \mathcal{M}(\gamma_{1,2})&=&\mathcal{R}(\gamma_{1,2})\mathcal{M}\mathcal{R}^T(\gamma_{1,2}), \\
    \mathcal{A}(\gamma_A)&=&\mathcal{R}(\gamma_A)\mathcal{A}\mathcal{R}^T(\gamma_A),\\
    \mathcal{P}(\gamma_P)&=&\mathcal{R}(\gamma_P)\mathcal{P}\mathcal{R}^T(\gamma_P),
\end{array}
\end{equation}
where
\begin{equation}
    \mathcal{R}(\gamma_{i}) = \begin{pmatrix}
\cos \gamma_{i} & -\sin \gamma_{i}  \\
\sin \gamma_{i} & \cos \gamma_{i} 
\end{pmatrix},
\end{equation}
are the rotation matrices on angles $\gamma_{i}$, where $i=1,2,A,P$. The outgoing electric field takes the form
\begin{equation}
\begin{array}{rcl}
    \mathbf{E}_{\parallel}(\gamma_1,\gamma_2,\gamma_A,\gamma_P) & = &\mathcal{A}(\gamma_A) \mathcal{M} (\gamma_2) \mathcal{M} (\gamma_1) \mathcal{P}(\gamma_P) \mathbf{E}_0, \\
    \mathbf{E}_{\perp}(\gamma_A,\gamma_P) & = &\mathcal{A}(\gamma_A) \mathcal{M}_{\perp}  \mathcal{M}  \mathcal{P}(\gamma_P) \mathbf{E}_0
\end{array}
\end{equation}   
and intensities in this case are $I_{\parallel}(\gamma_1,\gamma_2,\gamma_A,\gamma_P)=|\mathbf{E}_{\parallel}(\gamma_1,\gamma_2,\gamma_A,\gamma_P)|^2$ and $I_{\perp}(\gamma_A,\gamma_P)=|\mathbf{E}_{\perp}(\gamma_A,\gamma_P)|^2$.

By analogy with the ideal case, we can obtain the expression for the approximation of the deviated $V$
\begin{equation}
\begin{array}{rcl}
    V  & = & 1 - 4\alpha + 10\alpha^2 - \dfrac{1}{2}\beta^2 + 2\beta(\Gamma - \Xi) \\
     & &  - 2(\Gamma - \Xi)^2 + O_3(\alpha,\beta,\gamma_1,\gamma_2,\gamma_A,\gamma_P),
\end{array}
\end{equation}
where $\Gamma=\gamma_1+\gamma_2$ and $\Xi=\gamma_A+\gamma_P$. As we can see, in the second order  $V$ is most sensitive to the sums of these angles, provided that $T_s$ deviates from $T_p$: the main term is $2\beta(\Gamma-\Xi)$. See the expansion of the power series Eq.~(S3) in the {\bf \underline{supplementary material}}. The blue curve in Fig.~\ref{fig:fig3} represents the FMM-calculated $V(\gamma_1,\gamma_2,\gamma_A,\gamma_P)$ with  $\gamma_1 = -0.0175$ (i.e., $=-1^\circ$), $ \gamma_2=-0.0524$ ($=-3^\circ$), $\gamma_A = 0.0436$ ($=2.5^\circ$) and $ \gamma_P=0.0262$ ($=1.5^\circ$). It fits the measured $V_{\text{exp}}$ well over almost the entire frequency range of 250-650 GHz.

To conclude, we showed that the visibility function in the polarization contrast scheme and the fourth power of the normalized third Stokes parameter of the non-ideal QWP exhibit the same frequency dependence over a wide spectral range around the maximum values of the visibility function. Our analysis using the Fourier modal method, Jones matrix formalism, and analytical decomposition includes accounting for the transmission anisotropy, phase deviation, and misalignment of optical components from their ideal orientations. The established correspondence between these two quantities facilitates easy conversion between them, which can be practically exploited for rapid, broadband evaluation of the QWP's optical quality.

\section*{supplementary material}
See the supplementary material for additional analytical derivations and calculations of the visibility function and normalized third Stokes parameter, including the effects of angular misalignment of optical components in polarization contrast scheme.

\begin{acknowledgments}
The work was supported by the Russian Science Foundation (project No. 22-12-00351-$\Pi$).
\end{acknowledgments}

\section*{Data Availability Statement}
The data that support the findings of this study are available from the corresponding author upon reasonable request.

\FloatBarrier

\begin{thebibliography}{10}%
\makeatletter
\providecommand \@ifxundefined [1]{%
 \@ifx{#1\undefined}
}%
\providecommand \@ifnum [1]{%
 \ifnum #1\expandafter \@firstoftwo
 \else \expandafter \@secondoftwo
 \fi
}%
\providecommand \@ifx [1]{%
 \ifx #1\expandafter \@firstoftwo
 \else \expandafter \@secondoftwo
 \fi
}%
\providecommand \natexlab [1]{#1}%
\providecommand \enquote  [1]{``#1''}%
\providecommand \bibnamefont  [1]{#1}%
\providecommand \bibfnamefont [1]{#1}%
\providecommand \citenamefont [1]{#1}%
\providecommand \href@noop [0]{\@secondoftwo}%
\providecommand \href [0]{\begingroup \@sanitize@url \@href}%
\providecommand \@href[1]{\@@startlink{#1}\@@href}%
\providecommand \@@href[1]{\endgroup#1\@@endlink}%
\providecommand \@sanitize@url [0]{\catcode `\\12\catcode `\$12\catcode `\&12\catcode `\#12\catcode `\^12\catcode `\_12\catcode `\%12\relax}%
\providecommand \@@startlink[1]{}%
\providecommand \@@endlink[0]{}%
\providecommand \url  [0]{\begingroup\@sanitize@url \@url }%
\providecommand \@url [1]{\endgroup\@href {#1}{\urlprefix }}%
\providecommand \urlprefix  [0]{URL }%
\providecommand \Eprint [0]{\href }%
\providecommand \doibase [0]{http://dx.doi.org/}%
\providecommand \selectlanguage [0]{\@gobble}%
\providecommand \bibinfo  [0]{\@secondoftwo}%
\providecommand \bibfield  [0]{\@secondoftwo}%
\providecommand \translation [1]{[#1]}%
\providecommand \BibitemOpen [0]{}%
\providecommand \bibitemStop [0]{}%
\providecommand \bibitemNoStop [0]{.\EOS\space}%
\providecommand \EOS [0]{\spacefactor3000\relax}%
\providecommand \BibitemShut  [1]{\csname bibitem#1\endcsname}%
\let\auto@bib@innerbib\@empty
\bibitem [{\citenamefont {Grischkowsky}\ \emph {et~al.}(1990)\citenamefont {Grischkowsky}, \citenamefont {Keiding}, \citenamefont {van Exter},\ and\ \citenamefont {Fattinger}}]{Grischkowsky1990}%
  \BibitemOpen
  \bibfield  {author} {\bibinfo {author} {\bibfnamefont {D.}~\bibnamefont {Grischkowsky}}, \bibinfo {author} {\bibfnamefont {S.}~\bibnamefont {Keiding}}, \bibinfo {author} {\bibfnamefont {M.}~\bibnamefont {van Exter}}, \ and\ \bibinfo {author} {\bibfnamefont {C.}~\bibnamefont {Fattinger}},\ }\bibfield  {title} {\enquote {\bibinfo {title} {Far-infrared time-domain spectroscopy with terahertz beams of dielectrics and semiconductors},}\ }\href {\doibase 10.1364/JOSAB.7.002006} {\bibfield  {journal} {\bibinfo  {journal} {Journal of the Optical Society of America B}\ }\textbf {\bibinfo {volume} {7}},\ \bibinfo {pages} {2006--2015} (\bibinfo {year} {1990})}\BibitemShut {NoStop}%
  \bibitem [{\citenamefont {Jepsen}, \citenamefont {Cooke},\ and\ \citenamefont {Koch}(2011)}]{Jepsen2011}%
  \BibitemOpen
  \bibfield  {author} {\bibinfo {author} {\bibfnamefont {P.~U.}\ \bibnamefont {Jepsen}}, \bibinfo {author} {\bibfnamefont {D.~G.}\ \bibnamefont {Cooke}}, \ and\ \bibinfo {author} {\bibfnamefont {M.}~\bibnamefont {Koch}},\ }\bibfield  {title} {\enquote {\bibinfo {title} {Terahertz spectroscopy and imaging---modern techniques and applications},}\ }\href {\doibase 10.1002/lpor.201000002} {\bibfield  {journal} {\bibinfo  {journal} {Laser \& Photonics Reviews}\ }\textbf {\bibinfo {volume} {5}},\ \bibinfo {pages} {124--166} (\bibinfo {year} {2011})}\BibitemShut {NoStop}%
\bibitem [{\citenamefont {Katletz}\ \emph {et~al.}(2012)
\citenamefont {Katletz},
\citenamefont {Pfleger}, 
\citenamefont {Puhringer},
\citenamefont {Mikulics}, 
\citenamefont {Vieweg},
\citenamefont {Peters}, 
\citenamefont {Scherger},
\citenamefont {Scheller}, 
\citenamefont {Koch},
\ and\ \citenamefont {Wiesauer}}]
{Katletz2012}%
  \BibitemOpen
  \bibfield  {author} 
  {\bibinfo {author} 
  {\bibfnamefont {S.}~\bibnamefont {Katletz}}, 
  \bibinfo {author} 
  {\bibfnamefont {M.}~\bibnamefont {Pfleger}}, 
  \bibinfo {author} 
  {\bibfnamefont {H.}~\bibnamefont {Puhringer}}, 
   \bibinfo {author} 
 {\bibfnamefont {M.}~\bibnamefont {Mikulics}}, 
  \bibinfo {author} 
  {\bibfnamefont {N.}~\bibnamefont {Vieweg}}, 
  \bibinfo {author} 
  {\bibfnamefont { O.}~\bibnamefont {Peters}},
  \bibinfo {author} 
  {\bibfnamefont { B.}~\bibnamefont {Scherger}}, 
  \bibinfo {author} 
  {\bibfnamefont { M.}~\bibnamefont {Scheller}},
    \bibinfo {author} 
  {\bibfnamefont { M.}~\bibnamefont {Koch}},
  \ and\ 
  \bibinfo {author} 
  {\bibfnamefont {K.}~\bibnamefont {Wiesauer}},\ }
  \bibfield  {title} 
  {\enquote {\bibinfo {title} {Polarization sensitive terahertz imaging: detection of birefringence and optical axis},}\ }
  \href {\doibase 10.1364/OE.20.023025} 
  {\bibfield  {journal} {\bibinfo  {journal} 
  {Optics Express}\ }
  \textbf {\bibinfo {volume} {20}},\ \bibinfo {pages} {23025--23035} 
  (\bibinfo {year} {2012})}
  \BibitemShut {NoStop}%
\bibitem [{\citenamefont {Choi}\ \emph {et~al.}(2019)
\citenamefont {Cheng},
\citenamefont {Cheng}, 
\citenamefont {Huang},
\citenamefont {Zhang}, 
\citenamefont {Norris},
\ and\ \citenamefont {Kotov}}]
{Choi2019}%
  \BibitemOpen
  \bibfield  {author} 
  {\bibinfo {author} 
  {\bibfnamefont {W. J.}~\bibnamefont {Choi}}, 
  \bibinfo {author} 
  {\bibfnamefont {G.}~\bibnamefont {Cheng}}, 
  \bibinfo {author} 
  {\bibfnamefont {Z.}~\bibnamefont {Huang}}, 
   \bibinfo {author} 
 {\bibfnamefont {S.}~\bibnamefont {Zhang}}, 
  \bibinfo {author} 
  {\bibfnamefont {T. B.}~\bibnamefont {Norris}}, 
    \ and\ 
  \bibinfo {author} 
  {\bibfnamefont {N. A.}~\bibnamefont {Kotov}},\ }
  \bibfield  {title} 
  {\enquote {\bibinfo {title} {Terahertz circular dichroism spectroscopy of
biomaterials enabled by kirigami polarization
modulators},}\ }
  \href {\doibase 10.1038/s41563-019-0404-6} 
  {\bibfield  {journal} {\bibinfo  {journal} 
  {Nature Materials}\ }
  \textbf {\bibinfo {volume} {18}},\ \bibinfo {pages} {820–826} 
  (\bibinfo {year} {2019})}
  \BibitemShut {NoStop}%
\bibitem [{\citenamefont {Goldstein}(2011)}]{Goldstein2011}%
  \BibitemOpen
  \bibfield  {author} {\bibinfo {author} {\bibfnamefont {D.~H.}\ \bibnamefont {Goldstein}},\ }\href {\doibase 10.1201/b10436} {\emph {\bibinfo {title} {Polarized Light}}},\ \bibinfo {edition} {3rd}\ ed.\ (\bibinfo  {publisher} {CRC Press},\ \bibinfo {address} {Boca Raton},\ \bibinfo {year} {2011})\BibitemShut {NoStop}%
\bibitem [{\citenamefont {Collett}(2005)}]{Collett2005}%
  \BibitemOpen
  \bibfield  {author} {\bibinfo {author} {\bibfnamefont {E.}~\bibnamefont {Collett}},\ }\href {https://books.google.co.uk/books?id=5lJwcCsLbLsC} {\emph {\bibinfo {title} {Field Guide to Polarization}}},\ Field Guides\ (\bibinfo  {publisher} {SPIE Press},\ \bibinfo {year} {2005})\BibitemShut {NoStop}%
\bibitem [{\citenamefont {Zhang}\ and\ \citenamefont {Gong}(2015)}]{Zhang2015}%
  \BibitemOpen
  \bibfield  {author} {\bibinfo {author} {\bibfnamefont {B.}~\bibnamefont {Zhang}}\ and\ \bibinfo {author} {\bibfnamefont {Y.}~\bibnamefont {Gong}},\ }\bibfield  {title} {\enquote {\bibinfo {title} {Achromatic terahertz quarter waveplate based on silicon grating},}\ }\href {\doibase 10.1364/OE.23.014897} {\bibfield  {journal} {\bibinfo  {journal} {Optics Express}\ }\textbf {\bibinfo {volume} {23}},\ \bibinfo {pages} {14897--14902} (\bibinfo {year} {2015})}\BibitemShut {NoStop}%
\bibitem [{\citenamefont {Whittaker}\ and\ \citenamefont {Culshaw}(1999)}]{Whittaker1999}%
  \BibitemOpen
  \bibfield  {author} {\bibinfo {author} {\bibfnamefont {D.~M.}\ \bibnamefont {Whittaker}}\ and\ \bibinfo {author} {\bibfnamefont {I.~S.}\ \bibnamefont {Culshaw}},\ }\bibfield  {title} {\enquote {\bibinfo {title} {Scattering-matrix treatment of patterned multilayer photonic structures},}\ }\href {\doibase 10.1103/PhysRevB.60.2610} {\bibfield  {journal} {\bibinfo  {journal} {Phys. Rev. B}\ }\textbf {\bibinfo {volume} {60}},\ \bibinfo {pages} {2610--2618} (\bibinfo {year} {1999})}\BibitemShut {NoStop}%
\bibitem [{\citenamefont {Tikhodeev}\ \emph {et~al.}(2002)\citenamefont {Tikhodeev}, \citenamefont {Yablonskii}, \citenamefont {Muljarov}, \citenamefont {Gippius},\ and\ \citenamefont {Ishihara}}]{Tikhodeev2002}%
  \BibitemOpen
  \bibfield  {author} {\bibinfo {author} {\bibfnamefont {S.~G.}\ \bibnamefont {Tikhodeev}}, \bibinfo {author} {\bibfnamefont {A.~L.}\ \bibnamefont {Yablonskii}}, \bibinfo {author} {\bibfnamefont {E.~A.}\ \bibnamefont {Muljarov}}, \bibinfo {author} {\bibfnamefont {N.~A.}\ \bibnamefont {Gippius}}, \ and\ \bibinfo {author} {\bibfnamefont {T.}~\bibnamefont {Ishihara}},\ }\bibfield  {title} {\enquote {\bibinfo {title} {Quasiguided modes and optical properties of photonic crystal slabs},}\ }\href {\doibase 10.1103/PhysRevB.66.045102} {\bibfield  {journal} {\bibinfo  {journal} {Phys. Rev. B}\ }\textbf {\bibinfo {volume} {66}},\ \bibinfo {pages} {045102} (\bibinfo {year} {2002})}\BibitemShut {NoStop}%
\bibitem [{\citenamefont {Gippius}, \citenamefont {Tikhodeev},\ and\ \citenamefont {Ishihara}(2005)}]{Gippius2005}%
  \BibitemOpen
  \bibfield  {author} {\bibinfo {author} {\bibfnamefont {N.~A.}\ \bibnamefont {Gippius}}, \bibinfo {author} {\bibfnamefont {S.~G.}\ \bibnamefont {Tikhodeev}}, \ and\ \bibinfo {author} {\bibfnamefont {T.}~\bibnamefont {Ishihara}},\ }\bibfield  {title} {\enquote {\bibinfo {title} {Optical properties of photonic crystal slabs with an asymmetrical unit cell},}\ }\href {\doibase 10.1103/PhysRevB.72.045138} {\bibfield  {journal} {\bibinfo  {journal} {Phys. Rev. B}\ }\textbf {\bibinfo {volume} {72}},\ \bibinfo {pages} {045138} (\bibinfo {year} {2005})}\BibitemShut {NoStop}%
\end{thebibliography}

%
\providecommand{\noopsort}[1]{}\providecommand{\singleletter}[1]{#1}%
\end{document}


\title{Supplementary Material: \\ Visibility function and Stokes parameter analysis for non-ideal quarter-wave plates in a polarization contrast scheme}

\author{ S.\,A.~Kupavskii}

\affiliation{%
 Lomonosov Moscow State University, Faculty of Physics, Moscow 119991, Russia.
}%

\author{R.\,O.~Malikov}

\affiliation{%
 Lomonosov Moscow State University, Faculty of Physics, Moscow 119991, Russia.
}%
\affiliation{Prokhorov General Physics Institute of the Russian Academy of Sciences, Moscow 119991, Russia} 
\affiliation{Skolkovo Institute of Science and Technology, Moscow 143025, Russia}

\author{S.\,A.~Dyakov}%
\affiliation{Skolkovo Institute of Science and Technology, Moscow 143025, Russia}
\author{S.\,G.~Tikhodeev}

\affiliation{%
 Lomonosov Moscow State University, Faculty of Physics, Moscow 119991, Russia.
}%
\affiliation{Prokhorov General Physics Institute of the Russian Academy of Sciences, Moscow 119991, Russia} 
\affiliation{Skolkovo Institute of Science and Technology, Moscow 143025, Russia}

\author{A.\,A.~Gavrilov}
\affiliation{
Institute of Solid State Physics, Russian Academy of Sciences, Chernogolovka 142432, Moscow region, Russia
}

\author{K.\,R.~Dzhikirba}%
\affiliation{
Institute of Solid State Physics, Russian Academy of Sciences, Chernogolovka 142432, Moscow region, Russia
}

\author{S.\,A.~Andreeva}%
\affiliation{National Research University Higher School of Economics, Moscow 101000, Russia}
\affiliation{
Institute of Solid State Physics, Russian Academy of Sciences, Chernogolovka 142432, Moscow region, Russia
}

\author{I.\,V.~Kukushkin}%
\affiliation{
Institute of Solid State Physics, Russian Academy of Sciences, Chernogolovka 142432, Moscow region, Russia
}%

\author{N.\,A.~Gippius}%
\affiliation{Skolkovo Institute of Science and Technology, Moscow 143025, Russia}

\email{kupavskiysa@my.msu.ru}

\date{\today}

\maketitle

\section{The sixth order approximation in case of ideal QWPs alignment}

Considering an ideal QWPs alignment case, factoring up to the sixth and fifth order in small parameters $\alpha<1$ and $|\beta|<1$, respectively, is needed to reach the match between the FMM-calculated and approximated $V$ and $s_3^4$  almost in the whole range of frequency of 250-650 GHz, shown Fig.~2a of the main text:
\begin{equation}
\label{eq:V_asym_lossy_extended}
\begin{array}{ccc}
    V & = & 1 - 4\alpha + 10\alpha^2 - \dfrac{1}{2}\beta^2\ - 24 \alpha^3 + 2\alpha \beta^2 + \dfrac{1}{2}\beta^3      
    + 58 \alpha^4 - 7 \alpha^2 \beta^2 - 2\alpha \beta^3 - \dfrac{3}{8}\beta^4 - 140 \alpha^5\\
    & &   + 22\alpha^3 \beta^2 + 7\alpha^2\beta^3 +\dfrac{5}{4}\alpha \beta^4 + \dfrac{1}{4}\beta^5  
      +338 \alpha^6 -\dfrac{131}{2}\alpha^4 \beta^2 -22 \alpha^3\beta^3 \\
    & &   - \dfrac{29}{8} \alpha^2 \beta^4  
      - \dfrac{1}{2} \alpha \beta^5 - \dfrac{5}{32} \beta^6 + O_7(\alpha,\beta),
\end{array}
\end{equation}

\begin{equation}
\label{eq:Stokes_asym_lossy_extended}
\begin{array}{ccc}
    s_3^4 & = & 1 - 4\alpha + 6\alpha^2 - \dfrac{1}{2}\beta^2 - 4 \alpha^3 + 2\alpha \beta^2 + \dfrac{1}{2}\beta^3 
     + \alpha^4 - 3 \alpha^2 \beta^2 - 2\alpha \beta^3 - \dfrac{5}{16}\beta^4 \\
    &  & + 2\alpha^3 \beta^2 + 3\alpha^2 \beta^3 + \dfrac{5}{4} \alpha \beta^4 + \dfrac{1}{8}\beta^5 + O_6(\alpha,\beta).
\end{array}
\end{equation}
\begin{figure}[h]
    \includegraphics[width=0.8\columnwidth]{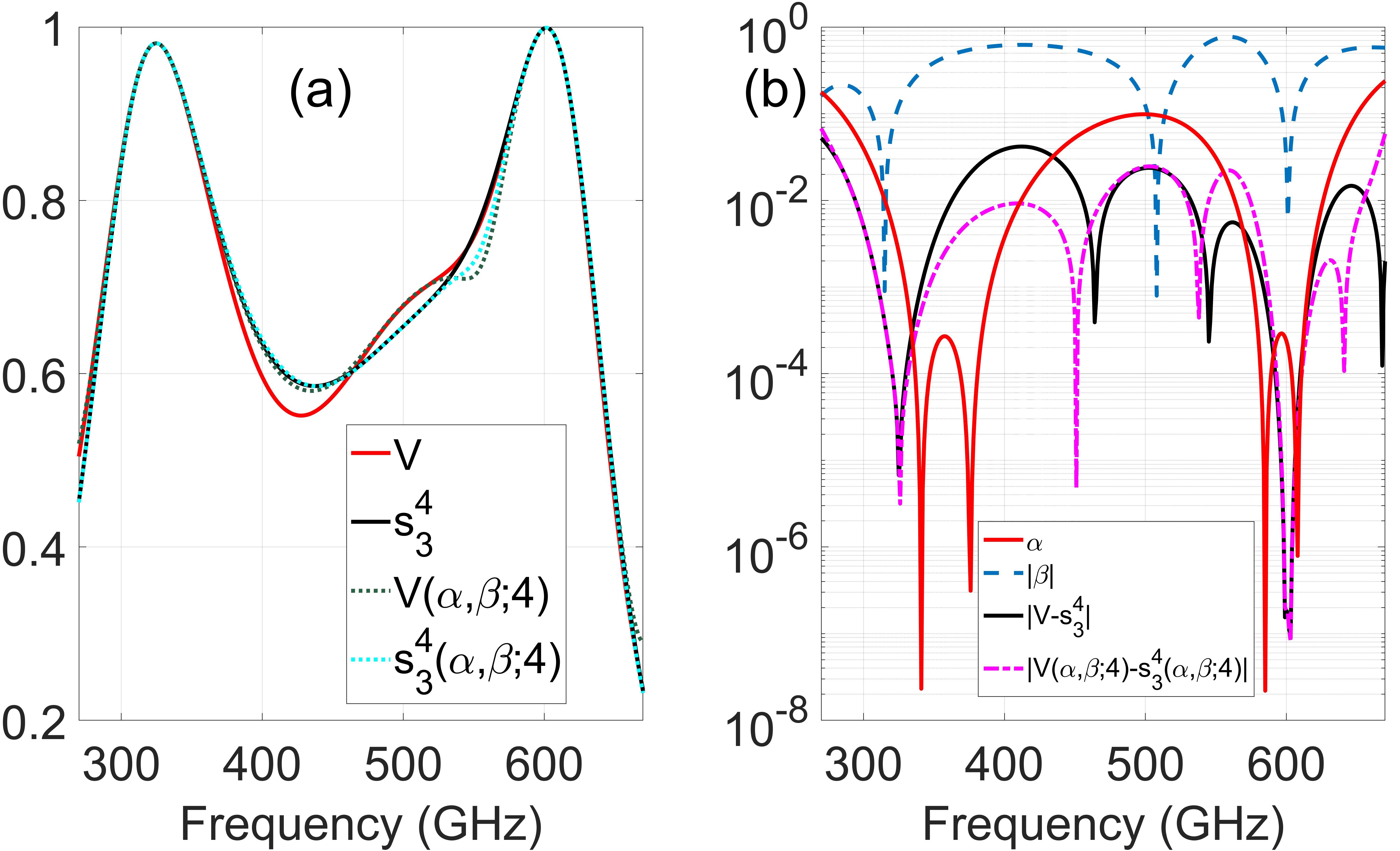}
    \caption{\label{fig:figS1}
    (a)~The FMM-calculated spectra of $V(\nu)$ and $s_3^4(\nu)$ and their fourth-order approximations in small parameters $\alpha(\nu)$ and $\beta(\nu)$ (Eqs.~(11) and (12) of the main text). (b) FMM-calculated $\alpha(\nu)$ and $|\beta(\nu)|$, as well as the absolute values of the differences between the FMM-calculated spectra of $V$ and $s_3^4$ and their fourth-order approximation Eqs.~(11) and (12) of the main text in a semi-logarithmic scale. 
}
 \end{figure}
Figure~\ref{fig:figS1}(a) shows that keeping the corrections up to fourth order only is not sufficient for the investigated here non-ideal QWPs. 
The frequency dependencies of small parameters $\alpha(\nu)$ and $|\beta(\nu)|$, calculated in accordance with FMM by Eqs.~(9) and (10) of the main text, respectively,
are shown in the
panel (b) in a semi-logarithmic scale.  
Note  the visually apparent difference between the FMM-calculated dependencies   and their fourth-order approximations.

\section{Accounting misalignment of all optical devices}

The approximation for the visibility function of the polarization contrast scheme in  small parameters $\alpha,\beta,\gamma_1,\gamma_2,\gamma_P,\gamma_A$ not higher than fifth power takes the form
\begin{equation}
\label{eq:V_asym_all_angle_deviation}
\begin{array}{ll}
    V & = 1 - 4\alpha + 10\alpha^2 - \dfrac{1}{2}\beta^2 + 2\beta(\Gamma - \Xi) - 2(\Gamma - \Xi)^2 - 4\sqrt{2\alpha}\,(\gamma^2 - \xi^2) \\
& - 24\alpha^3 + 2\alpha\beta^2 - 8\alpha\beta(\Gamma - \Xi) + 12\alpha(\Gamma - \Xi)^2 + 8\alpha(\gamma^2 - \xi^2) + \dfrac{1}{2}\beta^3 - \beta^2(\Gamma - \Xi) \\
& + 58\alpha^4 - 2\alpha\beta^3 - 7\alpha^2\beta^2 + 4\alpha\beta^2(\Gamma - \Xi) - \dfrac{3}{8}\beta^4 + \dfrac{5}{2}\beta^2(\Gamma - \Xi)^2 + \beta^2(\gamma^2 - \xi^2) - \dfrac{17}{3}\beta(\Gamma - \Xi)^3 \\
& - \dfrac{11}{4}\beta(\Gamma - \Xi)(\gamma^2 - \xi^2) + \dfrac{8}{3}(\Gamma - \Xi)^4 + \dfrac{3}{2}(\Gamma - \Xi)^2(\gamma^2 - \xi^2) - 2(\gamma^2 - \xi^2)^2 \\
& + 17\sqrt{2\alpha}\,\alpha(\gamma^2 - \xi^2) - 8\sqrt{2\alpha}\,\beta(\Gamma - \Xi)(\gamma^2 - \xi^2) + O_5(\alpha,\beta,\gamma_1,\gamma_2,\gamma_P,\gamma_A),
\end{array}
\end{equation}
where 
$\Gamma=\gamma_1+\gamma_2$, $\gamma=\gamma_1-\gamma_2$, $\Xi=\gamma_A+\gamma_P$, and $\xi=\gamma_A-\gamma_P$

As we can see, $\Gamma$ and $\Xi$ appear for the first time as a term  $2\beta(\Gamma-\Xi)$. As to $\gamma$ and $\xi$, they appear in even powers only, so we can replace them by their absolute values $|\gamma|$ and $|\xi|$, respectively. The lowest order with them is $-4\sqrt{2\alpha}(\gamma^2-\xi^2)$. 

\begin{figure}[t]
    \includegraphics[width=0.8\columnwidth]{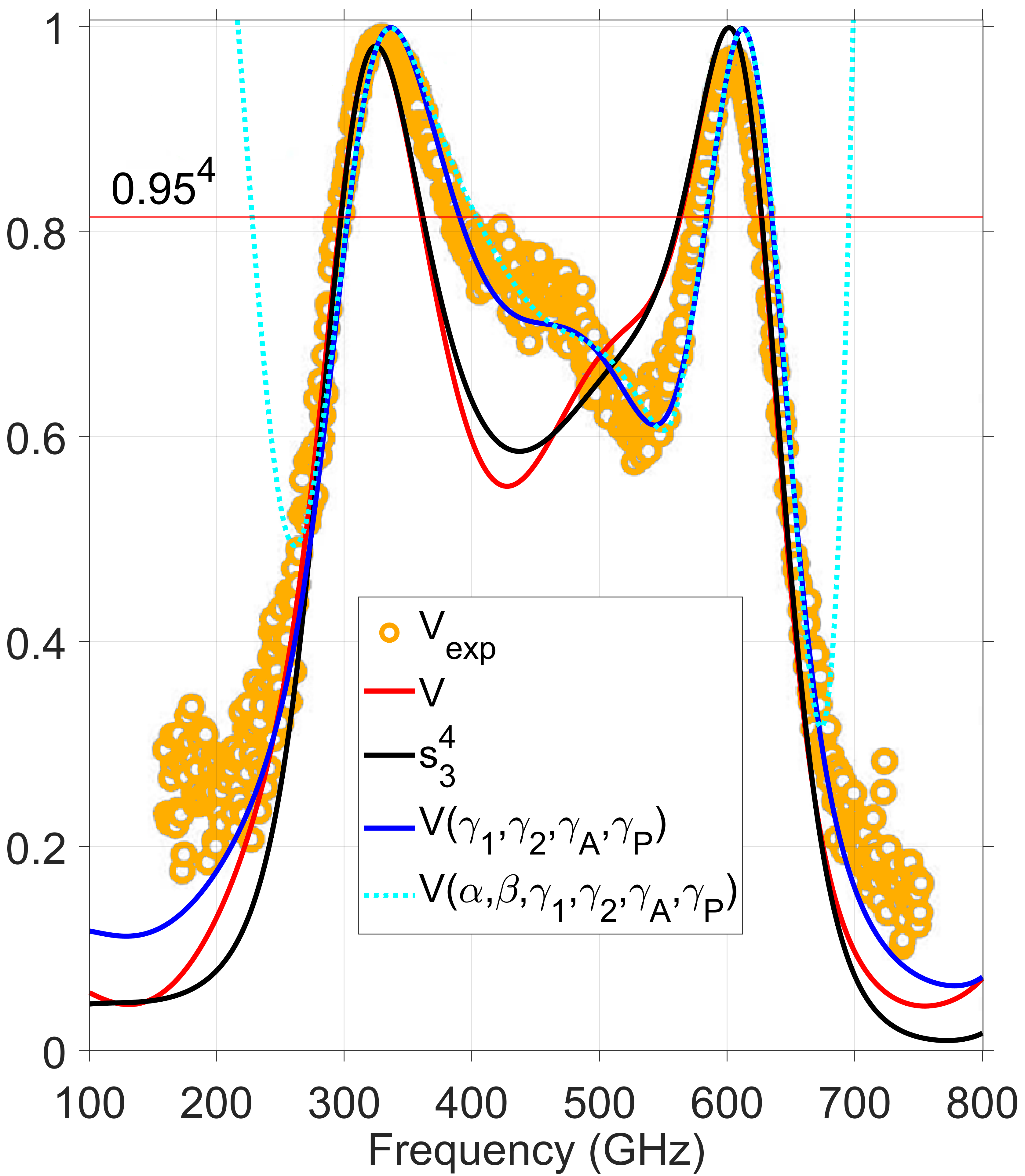}
    \caption{\label{fig:figS2} 
    The experimentally measured spectrum of the visibility function $V_{\text{exp}}$ (circles) for the polarization contrast scheme. The FMM-calculated visibility  $V$ (red curve) and $s_3^4$ (black) in the ideal alignment case. The FMM-calculated $V(\gamma_1,\gamma_2,\gamma_A,\gamma_P)$ (blue curve) in case of misalignment and its approximation by Eq.~(\ref{eq:V_asym_all_angle_deviation}) (light blue dotted curve) with $\gamma_1 = -0.0175$ (i.e., $=-1^\circ$), $ \gamma_2=-0.0524$ ($=-3^\circ$), $\gamma_A = 0.0436$ ($=2.5^\circ$) and $ \gamma_P=0.0262$ ($=1.5^\circ$). The red horizontal line indicates the  $0.95^4$ level.     
    }
\end{figure}

Figure~\ref{fig:figS2} compares the experimentally measured $V_{\text{exp}}$ with FMM-calculated $V$ and $s_3^4$ in the ideal alignment case as well as with  the  FMM-calculated $V(\gamma_1,\gamma_2,\gamma_A,\gamma_P)$ and its approximation by Eq.~(\ref{eq:V_asym_all_angle_deviation}) in the case of misalignment. As can be seen seen, the FMM-calculated  $V(\gamma_1,\gamma_2,\gamma_A,\gamma_P)$ with  $\gamma_1 = -0.0175$ (i.e., $=-1^\circ$), $ \gamma_2=-0.0524$ ($=-3^\circ$), $\gamma_A = 0.0436$ ($=2.5^\circ$) and $ \gamma_P=0.0262$ ($=1.5^\circ$) and its approximation by Eq.~(\ref{eq:V_asym_all_angle_deviation}) match around QWP's quality peaks frequencies. 